\documentclass{sigchi}
\usepackage[table]{xcolor}
\usepackage{color,soul}
\usepackage{caption}

\CopyrightYear{2020}
\setcopyright{rightsretained}
\doi{https://doi.org/10.1145/3313831.XXXXXXX}
\isbn{978-1-4503-6708-0/20/04}
\conferenceinfo{UIST'20,}{October  20--23, 2020, Minneapolis, MN, USA}
\acmPrice{\$15.00}

\toappear{\scriptsize Permission to make digital or hard copies of all or part of this work for personal or classroom use is granted without fee provided that copies are not made or distributed for profit or commercial advantage and that copies bear this notice and the full citation on the first page. Copyrights for components of this work owned by others than the author(s) must be honored. Abstracting with credit is permitted. To copy otherwise, or republish, to post on servers or to redistribute to lists, requires prior specific permission and/or a fee. Request permissions from permissions@acm.org. \\
{\emph{UIST '20, October 20--23, 2020, Virtual Event, USA} } \\
\copyright~2020 Association for Computing Machinery. \\
ACM ISBN 978-1-4503-7514-6/20/10\ ...\$15.00. \\
http://dx.doi.org/10.1145/3379337.3415842}

\usepackage{balance}       
\usepackage{graphicx}      
\usepackage{float}

\usepackage[T1]{fontenc}   
\usepackage{txfonts}
\usepackage{mathptmx}
\usepackage[pdflang={en-US},pdftex]{hyperref}
\usepackage{color}
\usepackage{booktabs}
\usepackage{textcomp}

\usepackage{microtype}        
\usepackage{ccicons}          

\usepackage{todonotes}

\usepackage{frame}
\usepackage{enumitem}
\usepackage{mathtools}
\usepackage{xspace}
\definecolor{lightblue}{rgb}{0.95,0.95,0.98}

\newcommand{\note}[2]{}

\newcommand{\ham}[1]{\note{orange}{Ham: #1}}

\renewcommand\hl[1]{#1} 

\def\plaintitle{mage: Fluid Moves Between Code and Graphical Work in Computational Notebooks}

\def\emptyauthor{}
\def\plainkeywords{Data Science Programming; Machine Learning Programming; Handoff; Computational Notebooks;}

\makeatletter
\def\url@leostyle{%
  \@ifundefined{selectfont}{
    \def\UrlFont{\sf}
  }{
    \def\UrlFont{\small\bf\ttfamily}
  }}
\makeatother
\def\pprw{8.5in}
\def\pprh{11in}

\definecolor{linkColor}{RGB}{6,125,233}
\hypersetup{%
  pdftitle={\plaintitle},
  pdfauthor={\emptyauthor},
  pdfkeywords={\plainkeywords},
  pdfdisplaydoctitle=true, 
  bookmarksnumbered,
  pdfstartview={FitH},
  colorlinks,
  citecolor=black,
  filecolor=black,
  linkcolor=black,
  urlcolor=linkColor,
  breaklinks=true,
  hypertexnames=false
}

\begin{document}

\title{\plaintitle}

\numberofauthors{6}
\author{%
  \alignauthor{Mary Beth Kery\\
    \affaddr{Carnegie Mellon University}\\
    \email{mkery@cs.cmu.edu}}\\
  \alignauthor{Donghao Ren\\
    \affaddr{Apple Inc.}\\
    \email{donghao@apple.com}}\\
  \alignauthor{Fred Hohman\\
    \affaddr{Georgia Institute of Technology}\\
    \email{fredhohman@gatech.edu}}\\
  \alignauthor{Dominik Moritz\\
    \affaddr{Apple Inc.}\\
    \email{domoritz@apple.com}}\\
  \alignauthor{Kanit Wongsuphasawat\\
    \affaddr{Apple Inc.}\\
    \email{kanitw@apple.com}}\\
  \alignauthor{Kayur Patel\\
    \affaddr{Apple Inc}\\
    \email{kayur@apple.com}}\\
}

\teaser{
  \vspace{-10pt}
  \includegraphics[width=\textwidth]{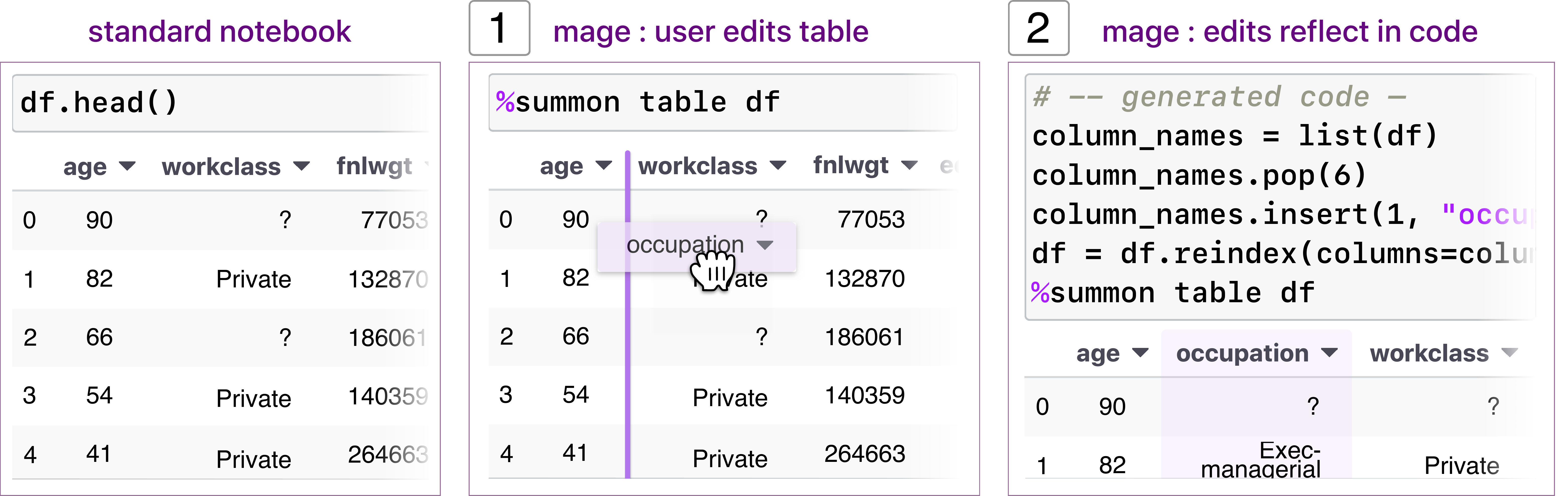}
  \caption{In a standard computational notebook, running code to refer to a data table \texttt{df} outputs a view-only rendering of the table. In a notebook powered by mage, the table is an interactive interface that the user can manipulate (1). When a user repositions a column in the interface (1), mage also automatically updates the code to reflect the change caused by the interaction (2).}
  \label{fig:teaser}
  }

\maketitle


\begin{abstract}
We aim to increase the flexibility at which a data worker can choose the right tool for the job, regardless of whether the tool is a code library or an interactive graphical user interface (GUI).
To achieve this flexibility, we extend computational notebooks with a new API mage, which supports tools that can represent themselves as both code and GUI as needed. We discuss the design of mage as well as design opportunities in the space of flexible code/GUI tools for data work. To understand tooling needs, we conduct a study with nine professional practitioners and elicit their feedback on mage and potential areas for flexible code/GUI tooling. We then implement six client tools for mage that illustrate the main themes of our study findings. Finally, we discuss open challenges in providing flexible code/GUI interactions for data workers.
\end{abstract}





\section{Introduction}
Data work is a complex domain where practitioners coming from diverse areas of expertise rely on a mix of tools to conduct their work~\cite{kdnuggets,kaggle:2017,kaggle:2018}.
Although code is the defacto power tool of choice for many professional data workers, graphical user interface (GUI) tools have an important role as well: charting tools, dashboards, and spreadsheets are all examples of pervasive forms of GUI tools for data work~\cite{kdnuggets,kaggle:2017,kaggle:2018}.
Graphical user interactions can provide easier inroads to novices, providing benefits such as recognition over recall~\cite{hutchins1985direct}, and are simply more practical for certain tasks, such as labeling image data.
In this paper, we aim to create an analysis environment where data workers have the flexibility to combine the complementary strengths of GUI and code and choose ``the right tool for the job''.


Computational notebooks make an ideal testbed for our research, as they are already set up for the purpose of blending mixed media with code.
A notebook is designed as a series of ``cells'' that form a single document: cells of code, cells of text, and cells of output of various formats~\cite{Kluyver:2016aa}.
This interactive document format has made notebooks highly popular.
Notebooks have been lauded by the scientific programming community for introducing new standards for replicability and understanding of data work --- because explanations, code, and output can be read and executed together in one place~\cite{Kluyver:2016aa, peng2011reproducible, perkel2018jupyter,shen2014interactive}.
However, aside from standard application menus and panes that live in the periphery of a user's notebook document, today's notebooks put strong restrictions on the kinds of GUI widgets supported within documents. There is no such thing as a ``GUI'' cell.
Today's community-created widgets are displayed in output cells, and include a variety of sliders, buttons, or visualizations that provide some, but very limited, forms of work.
More deeply blending programming with GUI-based work in notebooks presents technical and design challenges, but, we argue, is a natural area of growth for notebooks towards supporting full workflows in one place.

It should be noted that many recent examples from research and practitioner communities point to a desire to move notebooks and data computation in blended GUI/code directions.
Tools such as Wrangler~\cite{kandel2011wrangler}, Wrex~\cite{drososwrex}, Chartio, bamboolib, and qgrid\footnote{Spreadsheet widgets \href{https://bamboolib.8080labs.com}{bamboolib} or \href{https://github.com/quantopian/qgrid}{qgrid} both generate some form of code within notebooks.
Outside notebooks, Chartio has a \href{https://chartio.com/blog/why-we-made-sql-visual-and-how-we-finally-did-it/}{visual table approach to forming SQL queries} that also generates code.} all focus on providing flexible GUI or code editing for data transformation in tables.
As data tables is the canonical use case for so many of these tools, we demonstrate its appeal in \autoref{fig:teaser}.
In \autoref{fig:teaser} \ham{Better do 1.1,1.2,1.3, so this first reference isn't confusing}, the user sees a standard notebook computation where code \textit{output} is a preview of a data table.
As with most standard program output, notebook output is static and final, so the user is allowed to edit their table \textit{within code only}.
Alternatively, in \autoref{fig:teaser}.1 suppose the notebook \textit{did} offer GUI work in output. The user sees an interactive data sheet widget, where they can perform direct manipulation interactions as they would in spreadsheet software.
Simply putting an interactive data sheet into the notebook is an imperfect solution, however, since data sheets don't produce a visible record of actions like code does and for this reason are notoriously prone to user mistakes~\cite{powell2008critical}.
This limitation can be overcome by combining GUI with code: the user edits the data sheet widget via direct manipulation while their actions are \textit{simultaneously} reflected into auto-generated code (\autoref{fig:teaser}.2), providing a repeatable record.
Combining both GUI and code modalities gives users the complementary strengths of each as well as \textit{choice}.
The user is free to write code directly for some tasks or use direct manipulation to generate code for other tasks, as is most comfortable to them.

Our goal in this research is to design a generalized way to support a flexible GUI/code paradigm into a broad new class of data work tools.
Our contributions are:
\begin{enumerate}
  \item \textit{mage}: A generalized application programming interface (API) \textbf{mage} that can power tools to move fluidly between code and GUI in a notebook. \ham{Strictly, we enable developers to build new tools that allow users to move fluidly between code/GUI? (Tools don't really move, but users can move in the tools)}
  We discuss the design needs and tradeoffs for the API as well as how tool builders can integrate their work with mage.
  \item \textit{Demonstrating Design Space in Tools}: \ham{"Demonstrating" isn't a contribution. Demonstration is.}  We implement six example notebook widgets to support data workers on tasks from machine learning to visualization support, powered by mage. We use these tools to illustrate key themes in the kinds of toolings that flexible GUI/code ability enables. \ham{"GUI/Code" ability sounds a bit funny. "interplay between GUI and Code enables"}
  \item \textit{Practitioners Reflecting on the Design Space}: \ham{"Reflecting" is not a contribution. Our analysis of their reflection is.} To broaden our understanding of a flexible GUI/code tooling design space, we conduct an exploratory study with nine professional data works to gain their feedback and needs for the kinds of GUI/code tooling that might assist their daily work.
\end{enumerate}

Next, we discuss the API design of mage as grounded in prior work and needs specific to a general-use API.

\section{Prior Work \& Design Constraints}
Aside from data table transformation work mentioned~\cite{drososwrex, kandel2011wrangler}, prior work in flexible GUI/code work systems has focused on visual domains, such as drawing~\cite{hempel2019sketch}, UI design~\cite{hartmann2008design}, data visualizations~\cite{bigelow2016iterating}, and commercial 3D modeling software.
As these systems use various strategies for combining GUI with code, here we outline key design dimensions where mage draws upon, improves, or departs in key ways from prior work.

Allowing a user to make edits via GUI or code interchangeably in a live environment requires additional coordination in the underlying system.
As the user works, a system must be able to keep both code and GUI synchronized based on a shared underlying state.
Take for example a drawing app: if the user drew a star using a pen widget, the code for that drawing should update accordingly to represent a star.
If the user then wrote code to make the star blue, the star and widgets on screen should update accordingly to show the blue star.
Synchronizing between GUI actions and code edits becomes more complex the more power we give users to edit GUI and code interchangeably.
Why? User actions in GUIs can have ambiguous or multiple possible translations to code.
For instance, if a user pastes in a rotated star, is the star rotated positive 200 degrees or negative 160 degrees? User action in code come with the risk of there not being a GUI equivalent for that action. For instance, if the user writes mathematical constraints on their star shape, how should the GUI interface behave if it has no widget behavior built-in to represent constraints \cite{hempel2019sketch}?

For these reasons, common approaches carefully limit the expressiveness of what a user can do.
One such approach, which is built-in for notebooks today\footnote{ipyWidgets \url{https://github.com/jupyter-widgets/ipywidgets}}, is \textit{parameterization}.
Parameterization allows users to designate variables in code that they can then change at runtime through sliders and other widgets.
However, parameterization in the notebook today typically only changes the runtime value of a variable and does \textit{not} affect the value of the variable in written code\footnote{\hl{Note that parameterization widgets \textit{can change code} in some specific code editors, often for simple number or color variables. One implementations of notebooks that allows this is Carbide} \url{https://alpha.trycarbide.com}}.
This means that all GUI-based tuning only lasts for the current runtime session and is lost between sessions~\cite{hartmann2008design}.
By generating output code based on user interaction, mage can help parameterization widgets address this pitfall and better match a key design ethos of notebooks to \textit{support replicability}~\cite{Kluyver:2016aa}.



Beyond simple parameterization, another approach to balance between code expressivity and ambiguity is the use of a domain-specific language (DSL) where the creators carefully author mappings between a DSL and GUI actions~\cite{bigelow2016iterating,kandel2011wrangler,hempel2019sketch}.
Thus, for any action that the user takes, there is an equal representation of that action in both the code and GUI.
However, every DSL is \textit{by definition} designed for a specific task. \hl{Prime examples of DSLs in the data analysis space occur in commercial analysis platforms such SAS or JMP\footnote{\hl{SAS }\url{https://www.sas.com/en_us/home.html}\hl{ and JMP }\url{https://www.jmp.com/en_us/home.html}\hl{ are primarily GUI-based statistics and data analysis platforms that also support code in their own DSLs.}}. Both SAS and JMP are extensive GUI-based analysis tools fully compatible with code, but at the cost that all code must be in SAS's own or JMP's own bespoke DSL.
As everyday data work involves in a myriad of tasks in a variety of  programming languages,}
tying to just a single DSL conflicts with mage's goal as a general platform to enable a variety of tools.
As a result, the mage API does not directly use any DSL, but instead allows \textit{tool builders} to use DSLs
in their own tools where appropriate.
For example, our implementation of a plotting tool uses Vega-Lite~\cite{satyanarayan2016vega} as the underlying representation
for the charts.

Many systems have also used a programming by demonstration (PBD) approach to infer a correct mapping between what a user does in a GUI interface, and what the appropriate code to represent that should be~\cite{hempel2019sketch,drososwrex,kandel2011wrangler, gulwani2011automating}.
PBD has the benefit that, when carefully designed, it can ensure high-quality generated code synthesized fast enough to work in real time interactive applications~\cite{drososwrex,gulwani2011automating}.
Again, however, we find the generality goals of mage's API are at odds with this approach.
PBD, like domain specific languages, must be designed and tailored to a specific application, which is not feasible if mage must generalize to any data work application by any tool author. Again, a \textit{tool builder} could use PBD where appropriate \textit{internally} for their own tool, and their tool would be fully compatible to use with the mage API. The tension with generality is that the mage API itself cannot provide all the benefits of a PBD or DSL approach out-of-the-box, putting more effort on tool builders if they want to leverage these approaches.

As a departure from prior research, the goal of mage is not building a single highly flexible GUI/code tool,
but rather creating a platform for an ecosystem of tools.
Unlike a single closed system, like Sketch-n-Sketch for SVG editing~\cite{hempel2019sketch},
we cannot precisely model how a widget built with mage will affect state.
Sketch-n-Sketch contributed a \textit{complete and live 1-1 mapping} between GUI and code.
This mapping meant that code updates in realtime to reflect actions within the GUI, and conversely the GUI can update in realtime to reflect users' direct edits to the code.
This flexibility to interact with either code or GUI in realtime is what we hope tools built with mage can enable.
Next, we discuss how we design mage to achieve this goal, given the constraints that mage will not know precisely how any widget will behave.

\section{mage System Overview}
The goal of mage is an API that any tool builder can use to make their GUI widget flexibly communicate back and forth with user code and state in a computational notebook.
To facilitate this, mage is implemented as a Jupyter Notebook extension\footnote{Jupyter Notebook to date is the most popular open-source notebook platform, and offers extensive extension APIs \url{https://jupyter.org}}.
Since a Jupyter Notebook is itself a web app, mage is partly written in JavaScript to communicate with Jupyter APIs.
As our prototype system focuses on Python notebooks, some elements of mage are written in Python.

Next, we  use the \texttt{table} tool from \autoref{fig:teaser} as a use case to demonstrate how mage works, and how a tool creator can add bidirectional GUI/code communication to their widget:

\subsection{Getting Started}
\textbf{\texttt{table}:} Imagine the creator of \verb|table| wants to make an interactive spreadsheet-like tool that can be used to transform data in the notebook.
They start by designing and building a small web app for the table UI (we built \verb|table|'s UI as a single-page React app\footnote{Many of our example tools were created from a basic Create React App setup \url{https://create-react-app.dev}}).
The UI for the tool widget can be developed entirely outside the notebook at this phase, by using placeholder data (e.g.
a  cars sample dataset to display in \verb|table|) for what will later be data coming from the notebook user.

\textbf{mage:} The mage API first requires configuration.
A client tool provides its name, its parameter requirements for the data it would like to receive from the user, and its UI view as a JavaScript class.
When a user calls a client tool by its name, mage can create a new instance of the client tool's base JavaScript class and pass in all the parameters from the user.
Note that by asking tool creators to specify the exact number and types that their GUI tool will accept for parameters, mage is able to screen a request from the user and show users tool-specific error message if they've tried to call the client tool with data it cannot accept.

\textbf{\texttt{table}:} The tool creator registers ``\verb|table|'' as their tool's name, and the parameter requirement that the user pass a variable type that can be displayed in a table, such as a pandas dataframe\footnote{pandas is a common data manipulation library for Python optimized for performance with data tables \url{https://pandas.pydata.org}}.
The tool creator also specified its base JavaScript class is \verb|DataGrid| ---though, what exactly will \verb|DataGrid| do when it receives a Python binary object like a pandas dataframe? As an optional setting for situations like these, mage  will accept a Python function, \verb|preproccess|, which it runs on requested user data \textit{before} passing the data to the client tool.
The creator of \verb|table| authors their \verb|preprocess| function such that incoming data is converted into JSON objects that their system can read and display.

\textbf{mage:}  Now that mage has the configuration it needs, a user can call the \verb|table| tool by running ``\verb|%summon table df|'', where \verb|df| is the user's own data variable.
The figure below illustrates how the client tool appears for the user:

\begin{figure}[H]
    \centering
    \includegraphics[width=.85\columnwidth]{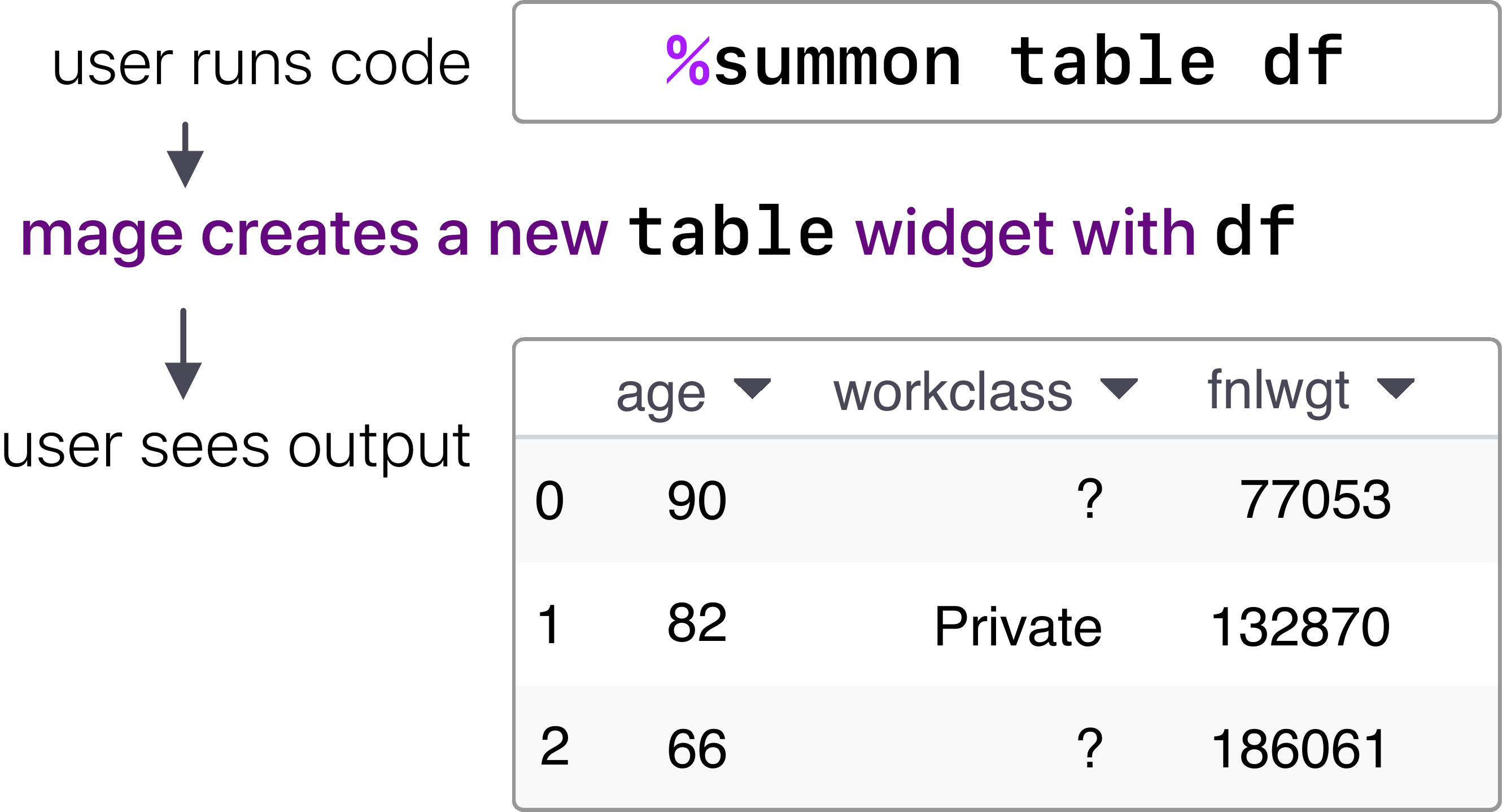}
    \vspace*{-0.1cm} 
    \label{mage-1}
\end{figure}

To keep with existing notebook conventions, a client tool in mage is instantiated from code and displayed below in that code cell's output.
mage leverages the ipywidgets library to render each tool's HTML div within a cell's output.

Users call mage via a custom \textit{magics} command syntax \verb|%summon <tool name> <parameters>|.
Magics are a keyword used in iPython that start with \verb|%|, and are conventionally used for meta-programming.
We chose magics to highlight to a user that the output produced by this code behaves differently from normal notebook output, but a different implementation of this work could easily pick a non-magic syntax.

\subsection{Choosing Which Actions Should Affect State}
Now that a client tool can be called, displayed, and run in the notebook, a tool creator could choose to stop here.
However, at this stage, even if the creator of \verb|table| builds all manner of table manipulation functionality into their UI, their UI alone won't be able to affect the user's data \verb|df| in any way.
Crucially, Jupyter notebooks use \textit{sandboxing} by design such that code that is not run by the user within their code cells cannot affect the user's program environment, and instead affects only a \textit{copy} of the user's runtime.
A user's variables cannot be directly modified without code run by the user.

Due to sandboxing, mage acts as the intermediary between a client tool and notebook state.
For a client tool's data preprocessing steps mage simply runs these within a sandbox, since a client tool should not be changing the user's variables at this stage.
Similarly, if a tool needs to request an updated value of a user's variable at runtime, mage provides a \verb|getVariable| call that will read and return (but not modify) user variable values.
For actions where the tool creator \textit{does} want to modify state, we describe mage's approach next.

\subsection{Templating Actions from GUI to Code}

\begin{figure}[t]
    \centering
    \includegraphics[width=.95\columnwidth]{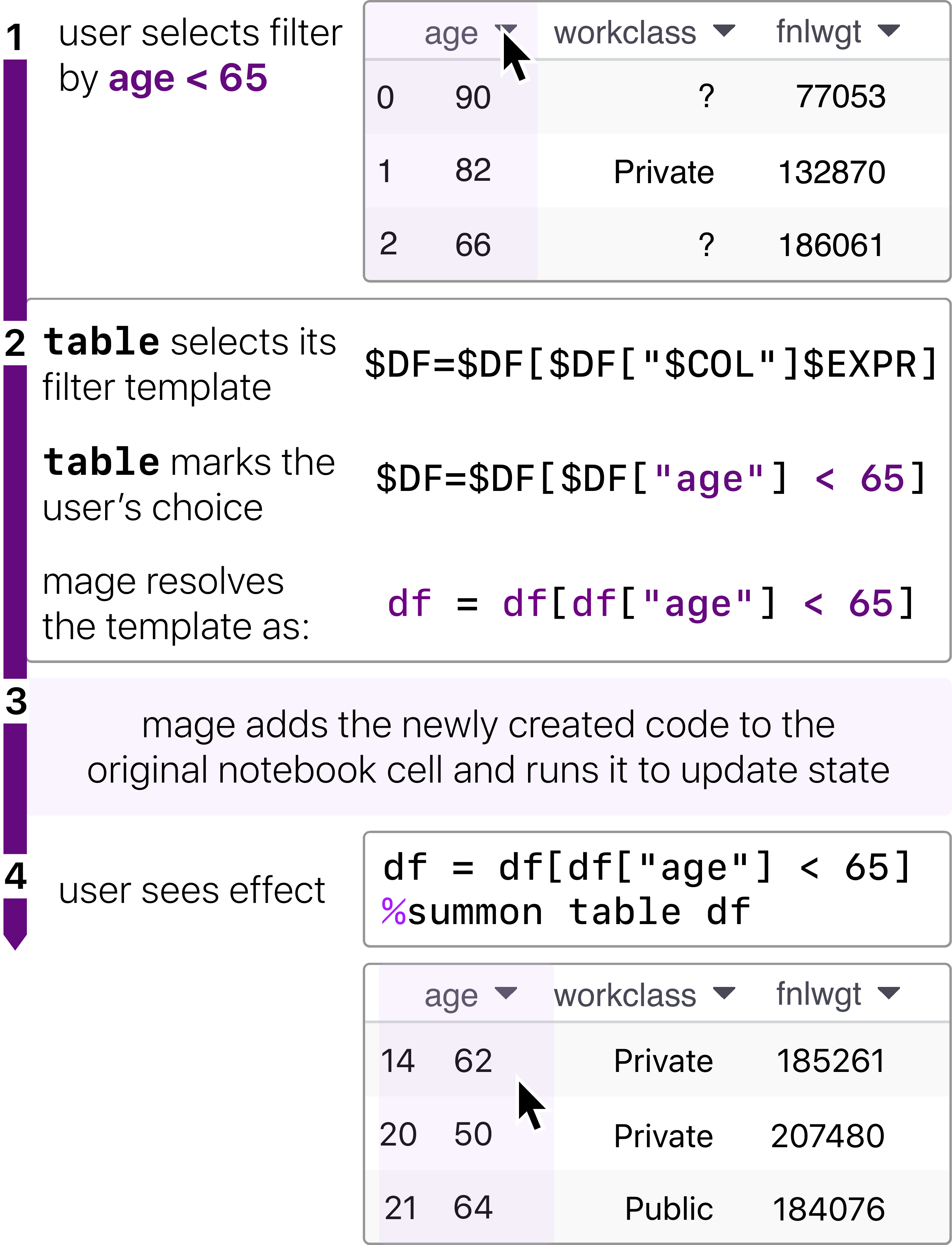}
    \caption{Process for reflecting actions in code in mage}
    \label{fig:mage-2}
\end{figure}

\textbf{\texttt{table}:} The creator of \verb|table| decides that when a user filters a column in the table, as illustrated in Figure~\ref{fig:mage-2}.1, they would like that filter applied to the user's data variable.
To implement this behavior, when the user starts a filter in the UI, the \verb|table| creator calls the mage API call \verb|handoff()|.

\textbf{mage:} To avoid notebook sandboxing, it's best to run code from the user's own code cells.
Thus, mage generates code needed to affect state and then injects it directly into the same cell where the client tool was originally instantiated.
However, mage does not inherently ``know'' what it means to filter a column of data, nor does it ``know'' the meaning of any other user action.
mage needs instructions on how to translate an action into code.
To provide the flexibility and expressivity necessary to support a broad range of tools, we chose to use \textit{code templates}.
For each action that should affect state, a tool author provides templates to reflect that action in code.

\textbf{\texttt{table}:} The tool creator writes a template such that when a user filters a column (\autoref{fig:mage-2}.1), \verb|table| has a fill-in-the-blank template corresponding to filtering a column using Python and pandas.
The format of a template is illustrated in \autoref{fig:mage-2}.2, where names that start with ``\$'' are blanks in the template and all other characters are literal pandas/Python syntax.
When the action occurs, \verb|table| calls the mage API \verb|handoff(<template>, <data>)|, with the template as well as data it ``knows'' about the user's action.
For instance, \verb|table| ``knows'' based on its UI that the user is filtering column ``age'' with the value ``age < 65'', so both of these values are passed along with the template as data to help fill in its blanks.

\textbf{mage:} Upon receiving a \verb|handoff()| API call, mage now resolves the template based on data the client tool provided as well as notebook state.
For instance, the blank \verb|$COL| in Figure~\ref{fig:mage-2}.2 is filled in with ``age'', and the blank \verb|$EXPR| is filled in with ``< 65''.
Since a client tool is \textit{not} responsible for knowing anything about the naming of variables in the user's environment, mage looks up the name of the user's data variable currently displayed in \verb|table| and fills in the blank \verb|$DF| with \verb|df|.
With all blanks in the template resolved, the code is complete Python syntax ready to run.
An extra benefit of templates is that a client tool can send mage \textit{multiple} templates for the same action, corresponding to the same action in different languages or library environments.
In this way, a tool creator can support a broad range of programming environment with relatively little extra effort.
If multiple template options are present, mage chooses the one that fits the current programming environment based on type checks.

Finally, mage inserts the new code into the code cell (\autoref{fig:mage-2}.3) and requests the notebook to re-execute the cell (\autoref{fig:mage-2}.4).
Re-executing the cell has several effects:
\begin{enumerate}
    \item Running the filter code ensures the user's variable \verb|df| is now filtered and will remain filtered for all other subsequent code anywhere in the notebook that uses \verb|df|.
    This filter is semi-permanent in that whenever the user starts a new session with their notebook or re-runs their notebook, the filter will execute on \verb|df| each time.
    To remove the filter, the user may delete that code using the normal notebook undo.\footnote{Note that we did not implement mage undo support for this prototype of the API, but it should readily follow by translating back code to action.
    By having mage watch the notebook undo stack it can notifying a client tool if their most recent template has been undone.}
    \item Running the cell refreshes the view of the tool \verb|table| and delivers the updated (filtered) data for \verb|df| to \verb|table| through its \verb|%summon| call parameter.
    The user sees a filtered table.
\end{enumerate}
For each subsequent action, the tool's templates for that action will be turned to new code by the same process.

\subsection{Reading Back Actions from Code to GUI}
\begin{figure}[t]
    \includegraphics[width=\columnwidth]{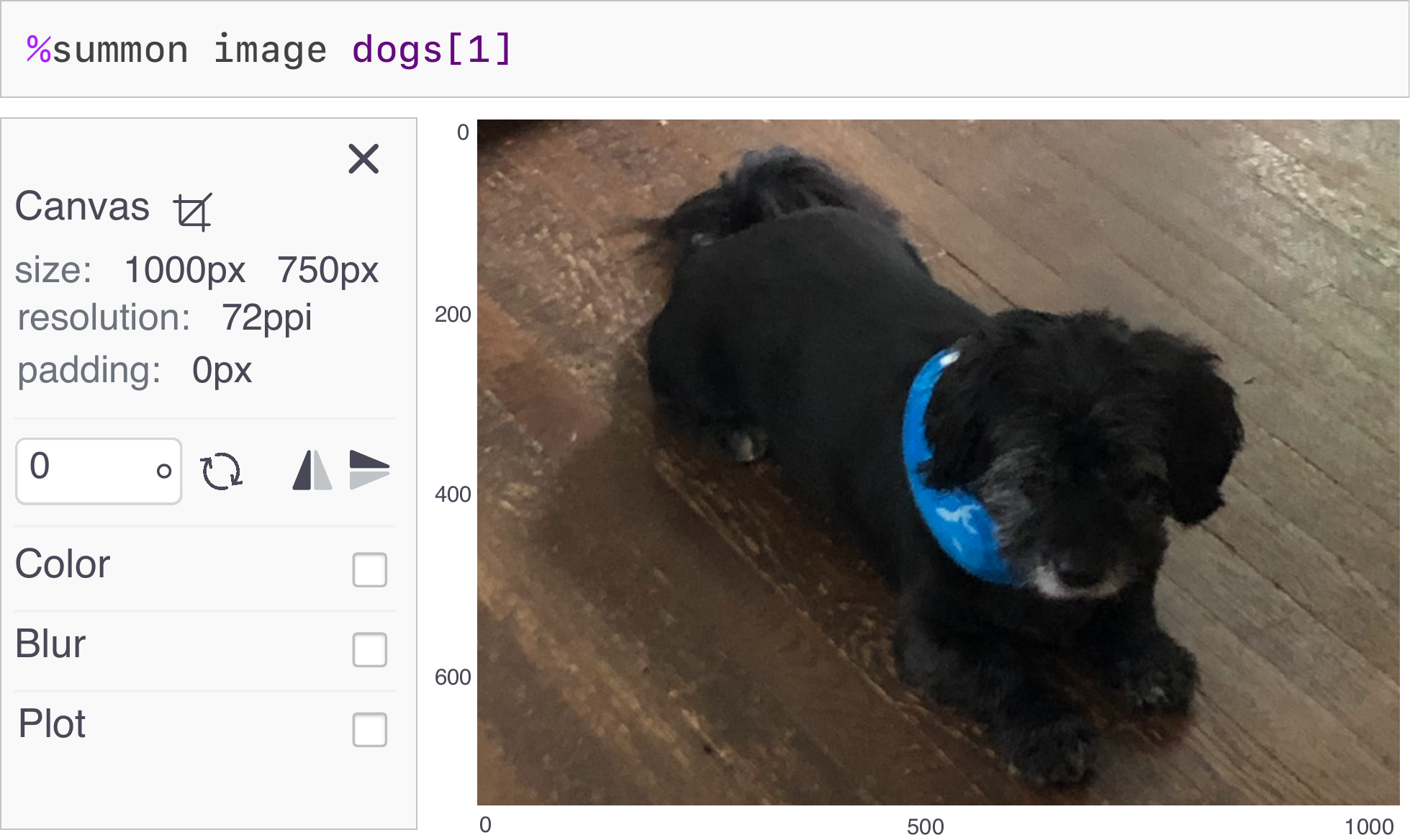}
    \caption{``image'' is a basic image editor using mage.
    Unlike writing plain code for image processing steps, the user can instantly see the effect of their operation on the image.}
    \label{image-editor}
\end{figure}

The goal of mage is to support \textit{any} GUI tool in a notebook to work fluidly with code.
However, many tools \textit{will not have a complete mapping between GUI action and code}.
To illustrate this challenge, we now switch our attention from \verb|table| to a new client tool, an image editor shown in \autoref{image-editor}.
Like most GUIs, the image editor has a finite set of features: it supports common image processing tasks, but not every image processing task.
Below, is the code that would be generated by mage if a user cropped their image with the \verb|image| tool:

\begin{figure}[H]
    \centering
    \includegraphics[width=.9\columnwidth]{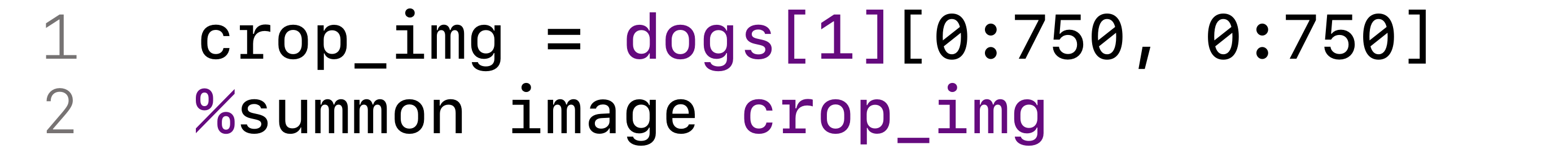}
    \vspace*{-0cm} 
    \label{code-image-1}
\end{figure}

Now suppose the user makes some edits to the code directly:

\begin{figure}[H]
    \centering
    \includegraphics[width=.9\columnwidth]{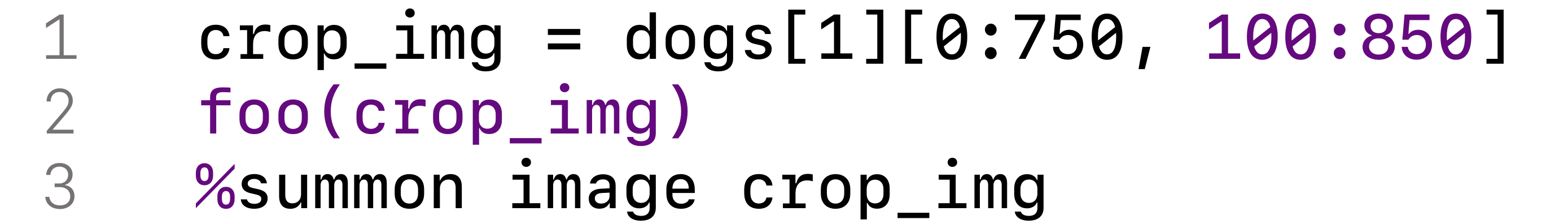}
    \vspace*{-0.3cm} 
    \label{code-image-1}
\end{figure}

In line 1, the user has changed the dimensions of the image crop.
mage watches code cells where a GUI tool is active and reacts to user code edits by attempting to update any state change back to the GUI tool.
Instead of writing actions, here we use a client tool's code template to \textit{read actions}: mage takes templates from \texttt{image} and fills in template blanks to create a regex that will ``recognize'' any code matching that template.
Here, the change to cropping is easily matched back to the crop action template.
Finally, mage reports the updated crop action back to the tool \texttt{image}.

Note that regular expressions are less powerful than other ways of recognizing code, such as using an abstract syntax tree (AST) analysis.
We choose to use regular expressions (in conjunction with type requirements provided by a template) as a conservative approach to limit recognizing false positives and to limit complications brought in by variables.
For instance, a user in line 1 could write ``\verb|n|'' instead of \texttt{850}, where the value of \verb|n| is \verb|850|.
A regex will reject ``\verb|n|'' for not being a valid number, while a more thorough static analysis could easily recognize ``\verb|n|'' as a variable representing a number.
However, recognizing \verb|n| comes at the cost of mage needing to watch the value of \verb|n| over time in case it changes from \verb|850| in the future.
Why? If \verb|image| is not notified as soon as the crop value is no longer \verb|850|, users may continue to use the image editor while seeing an invalid crop.
We defer these more complex state handling cases to future work, and for now avoid many of them with conservative approaches.

The new code the user has added to line 2 represents a different scenario.
The function \texttt{foo} is not within the \verb|image| GUI tool's limited action template list and mage has no inkling about how \texttt{foo} affects the image \texttt{crop\_img}.
Here mage responds by running the code, such that \texttt{image} displays the correct value of \texttt{crop\_img}, but with no ``understanding'' of any sequence of actions that precedes \texttt{foo}.
Since we cannot know if \texttt{foo} depends on the order of operations of the crop in line 1, mage takes a conservative approach and assumes all code preceding an unrecognized line of code is untouchable user code.
Thus \texttt{image} will have no undo operations possible for the image prior to \texttt{foo}, and all new actions will have auto-generated code inserted after \texttt{foo}.

\subsection{GUI Widgets with mage vs. GUI Notebook Extensions}
\hl{Due to the complexities of moving state between GUI/code representations, currently mage requires significant overhead from tool creators, in the form of code templates and type information. A creator might worry if flexible GUI/code interactions is too high a burden to implement for their tool. To address this limitation, we look more broadly at tool creators' current workflows to find other avenues to lower overhead.}

\hl{Today, aside from the simplest ipyWidgets which can be created by users on-the-fly, GUI widget tool creators for notebooks face much the same hurtles that tool creators face when extending any code editor. That is, creating a new widget is tightly tied to a specific editor's extension APIs. For instance, Jupyter Notebooks (used here for mage) has a different extension API from it's sibling notebook platform, Jupyter Lab. Given the still evolving nature of the notebook paradigm, today there are many notebook platforms to choose from, each with their own flavor of notebook programming and their own extension APIs. If we are to consider a GUI tools to be the same first-class means of work as code within a notebook, platform-dependency presents a major problem. A user can expect the same Python code to work the same across any notebook platform, but this will not hold true for GUI widgets.}

\hl{Our aim in mage is to make a tool creator's GUI widget as generalizable as possible to different platform and language environments. Thus, a tool creator with mage does \textit{not} take on platform-specific overhead, and provides only its own state-related logic: 1) an HTML-based GUI for their widget and 2) code templates to translate their GUI's interactions to user code. The mage API handles all notebook extension logic under-the-hood, which led to our implementation decision to build mage API on top of ipyWidgets API and the base notebook iPython runtime itself. The ipyWidgets API provides mage with some portability across the different notebook platforms that support ipyWidgets. Similarly, the base iPython runtime remains at the core of multiple different notebook platforms. Thus although mage API requires some modification to work in different notebook platforms, our core implementation is meant to be robust to different environments. Shielding individual GUI widgets from platform dependencies also comes with the benefit that changes to the underlying notebook extension APIs will only require update of the mage API and not individual implementation updates from tool creators.}

\section{Demonstrating Design Space in Tools}
At this point, we depart from discussing the mage API itself, to explore the design space of support tool interactions that a blended GUI/code environment powered by mage enables.

Thus far we have introduced two tools, \verb|table| and \verb|image| that leverage the mage API. Here, we describe all six tools, including the first two, that we implemented to explore different kinds of use cases for mage. To choose interesting use cases,  our team first brainstormed common data or machine learning tasks. Next, we iteratively sketched possible widgets for each task, and then narrowed down into the final set of six tools that represent the most diverse set of interaction ideas to push our understanding of the design space. From a research perspective, actively developing client tools pushes on our understanding of GUI/code interactions by allowing us to actively play with them, and also pushes on the kinds of scenarios our mage API should deal with and support.

\begin{figure}[t]
    \includegraphics[width=\columnwidth]{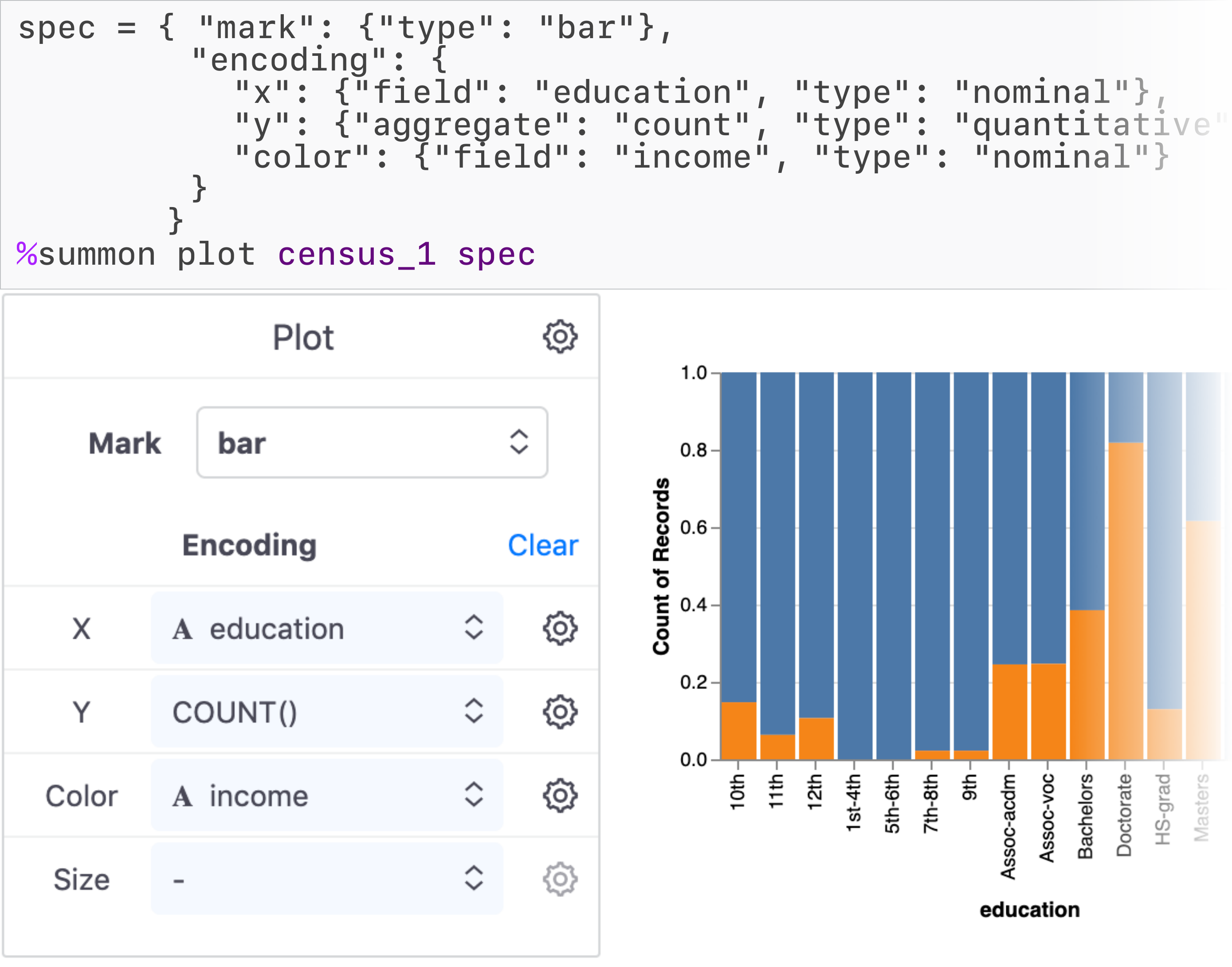}
    \caption{``plot'' is a chart builder tool powered by Vega-Lite.
    Instead of specifying upfront what kind of chart they would like, a user inputs just their data into \texttt{plot} and can then try out a variety of different chart types and encodings in the GUI.}
    \label{fig:plot}
\end{figure}

\subsection{\texttt{plot}: visualizing data as charts}

\verb|plot|, shown in \autoref{fig:plot}, is a chart builder that aims to help data workers iterate on common chart designs
without the need to recall an idiosyncratic chart syntax.
\verb|plot| uses a domain specific language (DSL), Vega-Lite~\cite{satyanarayan2016vega}, to represent state and to template its code.
Akin to Voyager~\cite{wongsuphasawat2015voyager} and Tableau~\cite{stolte2002polaris}, users can use \verb|plot|'s GUI to explore different visual encodings of their data.
Using mage's API for live code generation, \texttt{plot} generates a Vega-Lite~\cite{satyanarayan2016vega} specification based on user interaction.
Although our current implementation of \verb|plot| shows and reads Vega-Lite JSON specifications, future work can easily support bindings to Altair~\cite{VanderPlas2018}, the wrapper of Vega-Lite in Python.


As \verb|plot| aims to help users build common charts, it is intentionally less expressive than Vega-Lite, which supports a broad variety of charts, and thus supports only a small subset of Vega-Lite features.
One challenge with this expressivity mismatch is that the code and GUI state can be in conflict.
Similar to prior code/GUI tool like Adobe Dreamweaver\footnote{Adobe Dreamweaver blends code editing with GUI editing for website design \url{https://www.adobe.com/products/dreamweaver.html}},
\verb|plot| avoids such conflicts by ensuring that every Vega-Lite attribute either
has a widget for all of its possible values or have no interface for the attribute at all.
Basically, every widget in the GUI must be able to display any change to its corresponding attribute in the code.
For other Vega-Lite attributes without corresponding widgets, there are implicitly no conflicts
as the attributes are not shown in the GUI.



\subsection{\texttt{table}: transforming and cleaning data}
\verb|table|, as we have previously discussed and shown in \autoref{fig:teaser}, is a data sheet tool for transforming data through direct manipulation, much like common spreadsheet tools but without spreadsheet formulas. The goal of \verb|table| is to make data transformation and cleaning easier for data workers to do and to visually verify that their data looks correct during this work. So called ``data wrangling'' can be the most tedious and time-consuming part of a data worker's workflow~\cite{kandel2011wrangler}.

\subsection{\texttt{image}: editing image data}

\verb|image|, as introduced earlier in \autoref{image-editor}, is a basic image editor that uses mage to generate code for processing image. Since media data like images, sound, or video already have custom display widgets in a notebook today, here we explore the reasonable next steps of using mage to provide media editors standard to different kinds of media data. Although a data worker can certainly write code to process image data, the goal of \verb|image| is to give data workers a faster feedback loop to ``see'' the effects of their operations.


\subsection{\texttt{datasplit}: segmenting data for machine learning}

With the \verb|datasplit| tool, we explore the possibility of using mage to give a visual face to common data work API calls that contain lots of options. We designed \verb|datasplit| as a visual version of the common scikit-learn\footnote{scikit-learn is a common novice-friendly machine learning library in Python \url{https://scikit-learn.org/stable/}} function \verb|train_test_split()|. This means that \verb|datasplit| uses mage to write a call to \verb|train_test_split()|, while providing a GUI to give the user the ability to play and explore with all the parameter options for the API call. While \verb|train_test_split()| is one specific API call, the task it represents is a universal one for supervised machine learning: given a dataset, split the dataset into separate smaller datasets that can be used to train and test the model. We note that many library calls, especially in machine learning or data processing, are highly tunable with a large set of optional parameters. For a data worker, especially a novice one, the meaning and appropriate use of all these options can be hard to grasp~\cite{cai2019software}. Thus, the goal of \verb|datasplit| is to explore the idea of giving data workers more play-based interactions to understand options.
\ham{We should emphasize the reproductivity aspect here.  Unlike splitting with a traditional GUI, the generated code for data splitting can be re-run when they fix data cleaning commands upstream. (We should similarly see if we can add this aspect to other tools.)}

\begin{figure}[t]
    \includegraphics[width=\columnwidth]{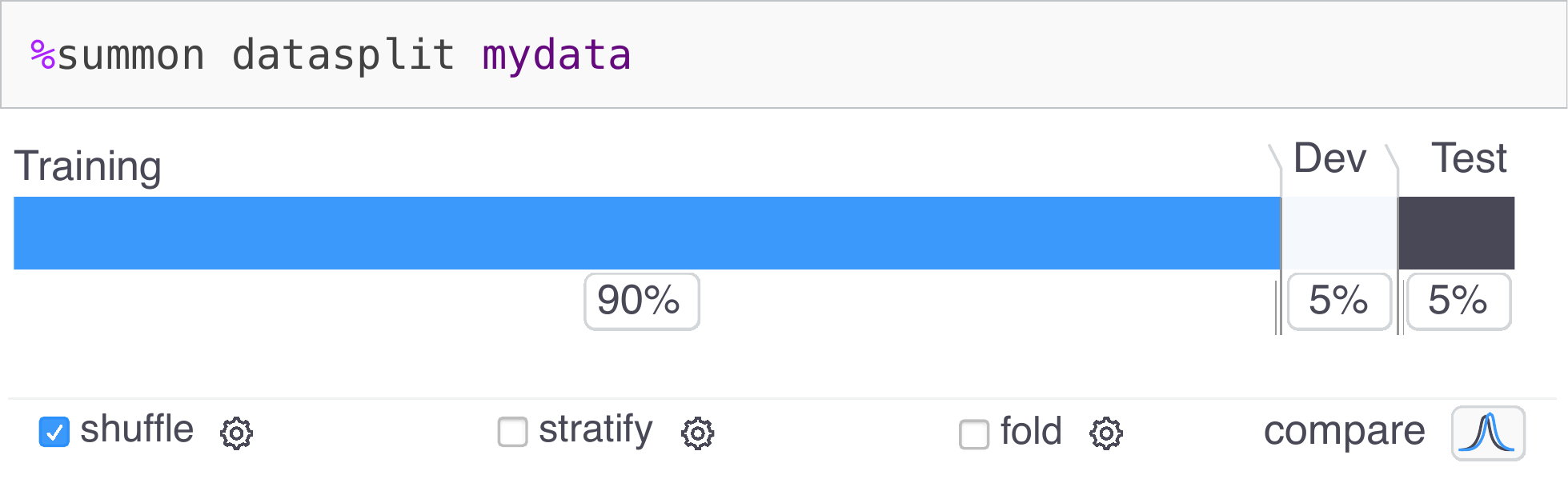}
    \caption{``datasplit'' is a tool for a standard operation of splitting data into subsets for training and testing a ML model.
    \texttt{datasplit} gives the user operations to explore choices around creating data subsets. The color-segmented bar visually offers the users a suggested default split, as well as tunable divides that allow them to explore different splits.}
    \label{fig:datasplit}
\end{figure}

\begin{figure}[tbh]
    \includegraphics[width=\columnwidth]{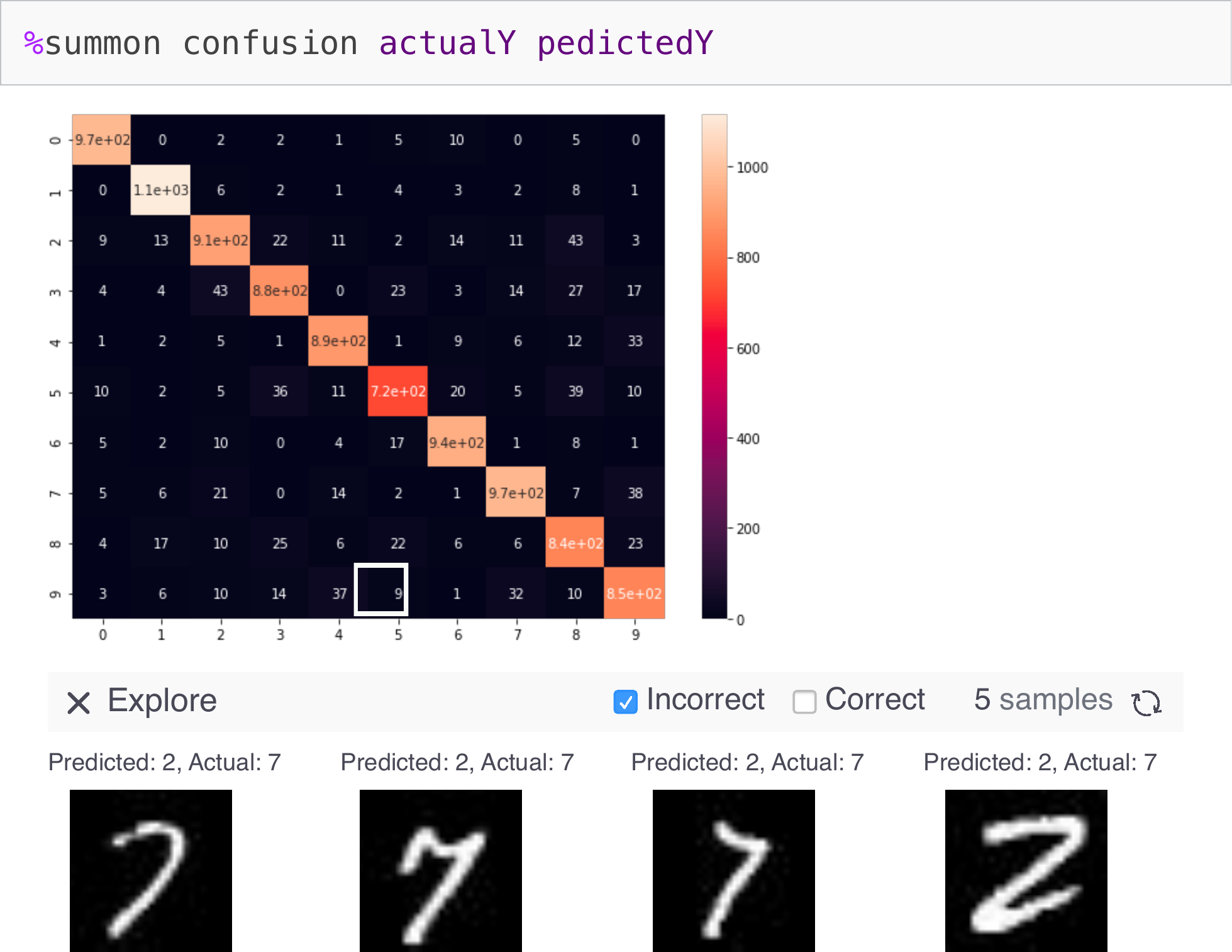}
    \caption{``confusion'' is a tool to explore the correct and incorrect results of a classifier model, using a classic confusion matrix visualization.
    Unlike a plain confusion matrix, the user can interactively explore each result in the matrix, which appears in a row of samples beneath the matrix.}
    \label{fig:confusion}
\end{figure}

\subsection{\texttt{confusion}: exploring classifier model performance}

\verb|confusion| is a tool, shown in \autoref{fig:confusion}, with the goal of helping data workers understand machine learning model performance in  their notebook.
A confusion matrix is a common visualization for the performance of a classifier.
The matrix shows each possible label on the x and y axis: a perfect performance is one were all predicted labels fall in the left-to-right diagonal.
Existing libraries like scikit-learn output a static table of numbers.
By contrast, in a mage powered tool, the matrix becomes an interactive exploration interface where the user can click on individual matrix cells to see below examples of data that were (mis)classified in that way. As compared to tools we've discussed so far, an interesting property of \verb|confusion| as an exploratory visualization is that its primary use of mage is not actually to generate code. Instead, \verb|confusion| uses the mage API to retrieve and process data from the user's environment to automatically generate different ways for the user to look at their classifier results in the UI, without asking the user to provide any additional code.


\begin{figure}[tbh]
    \includegraphics[width=\columnwidth]{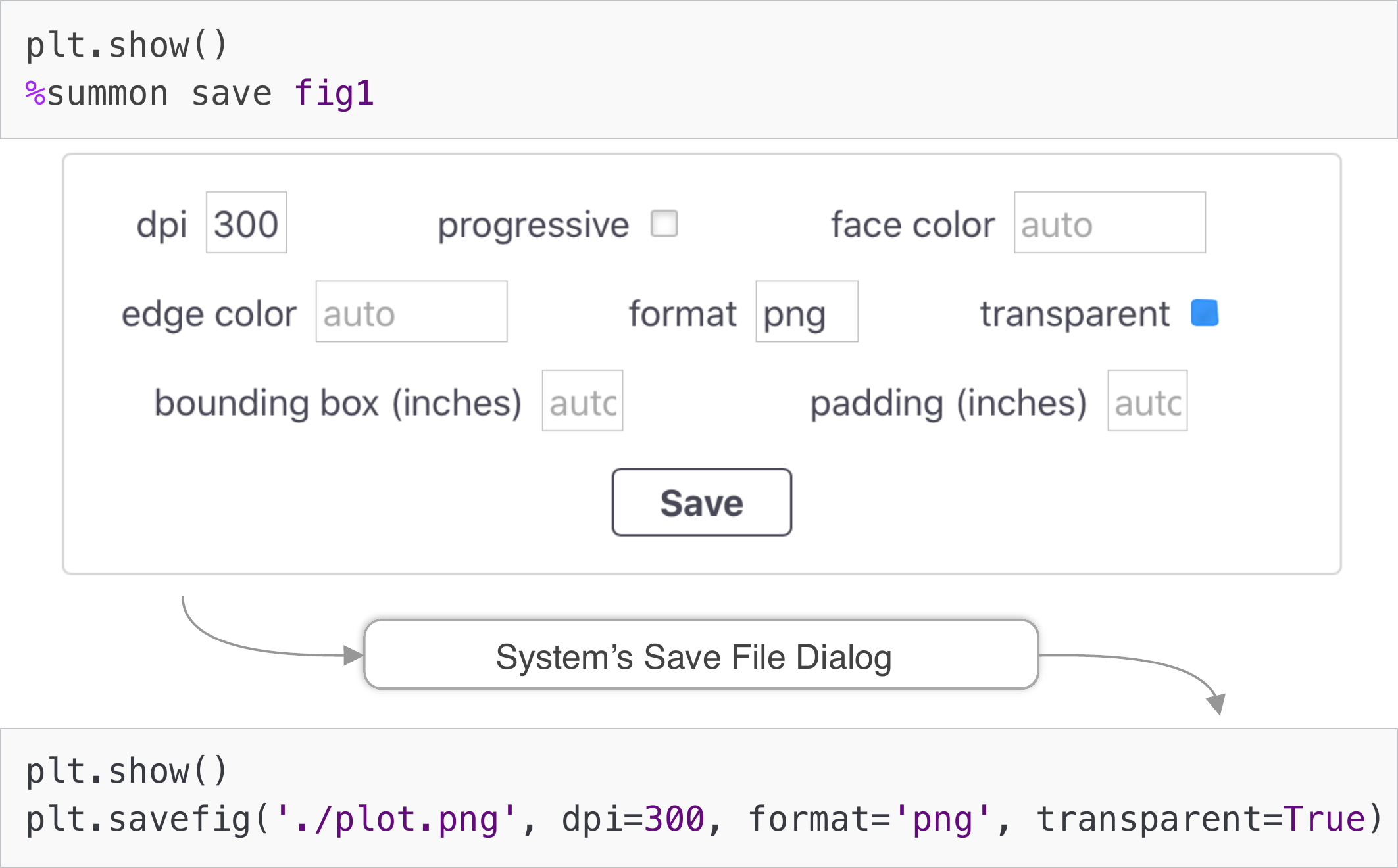}
    \caption{``save'' is a tool, similar to a standard save file dialogue box, that lets a user interactively choose between the options for saving a chart or figure.}
    \label{fig:io}
\end{figure}

\subsection{\texttt{save}: exporting (anything) to a file}
Similar to \verb|datasplit|, the \verb|save| tool puts a visual GUI face to library calls for exporting data to a file. Shown in Figure~\ref{fig:io}, when the user runs \verb|save| on a matplotlib plot, \verb|save| presents a dialogue box containing the various options for exporting the plot to file. Once the user clicks the ``save'' button, they can then use a standard system file browser to save the file to their desired location, while \verb|plot| generates appropriate saving code using mage. The key difference in idea for \verb|save| compared to \verb|datasplit|, is that we imagine \verb|save| could be used as a single visual GUI utility for \textit{many different} kinds of API calls that perform an export/save. Using mage to type-check the type of variable that a user inputs into the \verb|%summon save <data>| call, \verb|save| could contain bindings for different kinds of libraries, and adapt the dialogue box visualized accordingly. So, if the user inputs a pandas dataframe instead of a matplotlib plot, \verb|save| will offer the save options offered by the pandas library accordingly.


\begin{figure}[tbh]
    \includegraphics[width=\columnwidth]{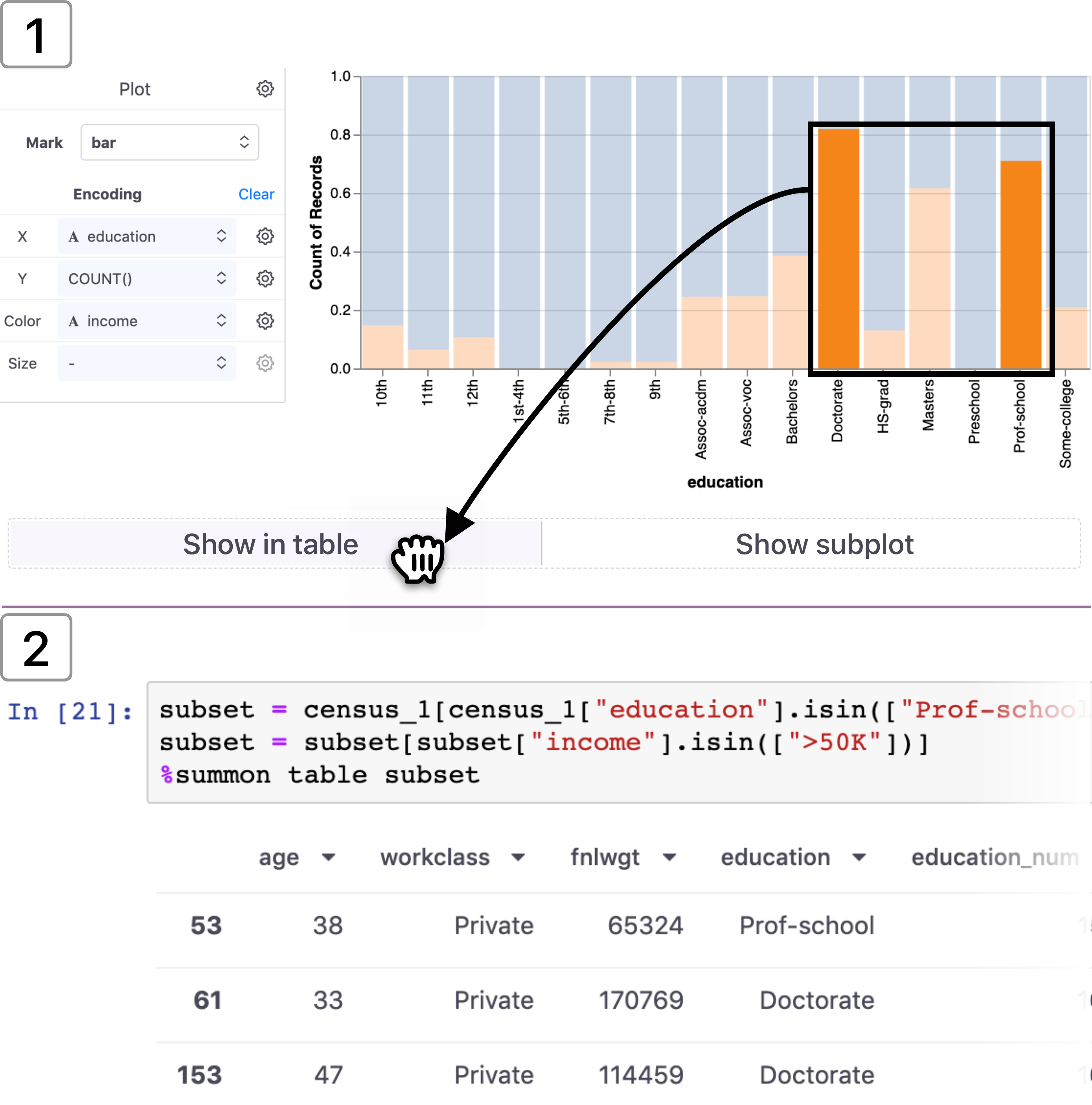}
    \caption{In (1) the user has the \texttt{plot} tool open and has constructed a chart.
    They select two columns from the chart of interest and begin dragging these columns outside of the chart.
    As soon as the user starts dragging, an option bin menu appears.
    The user drags and drops the columns to ``Show in table'', and in (2) a new cell is inserted immediately below that displays their selection both in code and in the \texttt{table} tool.}
    \label{fig:drag}
\end{figure}

\section{Multi-tool Usage Scenario}

Now that we have demonstrated six example client tools for mage, an important design consideration is how a user might use their notebook with \textit{multiple} flexible GUI/code tools working in concert. Although we largely leave the multi-tool design space for future work, we did design and implement one specific kind of interaction to illustrate the benefits fluid transitions between GUI/code tooling might offer users. We focused on one very fundamental data worker need: \textit{selecting data}.

Imagine an analyst is exploring the 1994 US Census dataset from the UCI Machine Learning Repository~\cite{UCI} and wants to investigate historical gender bias in high income and high education workers.
First, they call \verb|%summon plot census| to visualize the distribution of education as a bar chart in the \verb|plot| tool (\autoref{fig:drag}.1).
The analyst then adds ``income'' to the color channel
and enables ``normalize'' stacking for the ``Y'' axis to compare the percentage of low vs. high income in each education level (\autoref{fig:drag}.1).
The analyst sees that higher education is correlated with higher income.
Now, interested in the high education and high income group,
they select the ``>50k'' income bars with ``Doctorate'' and ``Prof school'' education levels, and begin dragging this selection. As they drag the data selection out of the plot, mage provides a ``drag bin'' interaction where options specific to the \verb|plot| client tool appear. The user sees two options for their selection: ``Show in table'' or ``Show in subplot'' (\autoref{fig:drag}.1). They drag their selection in to ``Show in table'', resulting in a new instance of the \verb|table| tool appearing in a new cell below, with that data slice already loaded (\autoref{fig:drag}.2).
In this new table view, the analyst finds that all five rows are ``Male'', leading to a hypothesis that there is a gender bias.

The interaction of dragging data selections between tools is the only multi-tool interaction currently built into the mage API. It can be used to drag a column from the \verb|table| tool to visualize distributions with the \verb|plot| tool or drag a matrix cell from the \verb|confusion| tool to see the full \verb|table| view of that particular data slice. Instead of GUI to GUI transitions, the user can also use drag to export a selection from a visualization like \verb|plot| into a plain code selection for that data slice. Our goal is to make it simpler and more flexible for data workers to move the same data or data slices between different modalities.

\begin{table}
  \renewcommand{\arraystretch}{1.2}
  \begin{center}
    \footnotesize
    \rowcolors{1}{}{lightblue}
    \begin{tabular}{ c | >{\arraybackslash}m{6cm} }
     {\small\bf Widget Name} & {\small\bf Common DS/ML Task} \\
     \hline
     \verb|table| & A spreadsheet widget that reflects direct manipulation actions in generated code.\\
     \verb|plot| & A chart that generates selection code when a user selects data through direct manipulation.\\
     \verb|image| & A basic image editor that generates code for common image data processing tasks. \\
     \verb|confusion| & An interactive tool for exploring classifier model performance.\\
     \verb|api| & A basic form interface for building correct calls to a web API to retrieve data.\\
     \verb|datasplit| & An interactive tool for splitting data into sets needed to train and test a model.\\
     \verb|save| & An interactive tool for saving a plot or figure with export settings. \\
    \hline
    \end{tabular}
    \caption{Initial paper prototypes shown to practitioners.}
    \label{table:ideas}
    \end{center}
\end{table}

\section{Practitioners Reflecting on the Design Space}
To better grasp practitioners' goals, interests, and values in this new design space of flexible GUI/code tooling, we recruited nine participants from a large technology company who both use notebooks and do professional data work. Most participants also worked with machine learning as part of their job.
Each session began with a brief pitch for a new system for GUI/code work in notebooks that our team was in early stages of developing.
We then asked participants to help our team understand which tool and interaction ideas would be helpful (or not) for their own daily work.

To ground the discussion and help illustrate what we were talking about, we presented participants with seven starter tool ideas (listed in \autoref{table:ideas}). Note that the \verb|api| tool has the same basic idea as the \verb|datasplit| tool in that they are both GUI tools to configure an API call, thus we did not actually implement the \verb|api| tool.

Each tool was shown as a high-fidelity paper prototype, except for \verb|table| and \verb|plot|.
These two were given as an initial example, chosen for their familiarity, and shown together within a live notebook. Participants were given the opportunity to try out the live \verb|table| and \verb|plot| tools with a simple data task to get a sense for the experience and feel of using a flexible GUI/code tool in the notebook.
Each subsequent tool was then shown one at a time on paper to focus discussion.

While it may come as a surprise that we chose to present most tools on paper when it was fully possible to keep this design session in a live notebook with functioning tools, we made this choice to better enable conceptual feedback participants. Conventional user-centered design wisdom as well as prior design research~\cite{davidoff2007rapidly} has found that when users are presented with high fidelity functioning prototypes, their feedback typically fixates on details of visuals and functionality of that particular implementation. In order to prevent distracting details from dominating the conversation and focus the user instead on considering the broader \texttt{concept} or ideas behind the tool, it is helpful to present less detail and lower fidelity for a tool to users~\cite{davidoff2007rapidly}. Thus, we kept much of the discussion to single-screenshot paper prototypes of tools on paper.

 When viewing each paper prototype, we asked participants to evaluate its ``usefulness'' as well as general impressions as if they were the one who would be using the tool.
 During this process, we encouraged participants to generate their own variants or entirely new tool ideas based on tasks they faced in their own daily work.
 We also encouraged participants to freely sketch or annotate directly on the paper prototypes.

\subsection{Study Limitations}
\hl{Note that by pushing participants to discuss more abstractly about concepts, rather than implementation details of mage or details of any of our demonstration tools, this study \textit{does not} serve as a usability evaluation of mage. At this stage in research, we considered it higher-priority to expand our understanding of the core needs mage and flexible GUI/code tooling \textit{should} support. An important next step of our future work is the evaluate the usability and efficacy of mage for serving these needs expressed by practitioners, which we detail next.}

\subsection{Results}
Following the study, we conducted a thematic analysis of participants' tool ideas and feedback, which interactions they valued or rejected and why.
We next discuss these themes and illustrate how these themes play out with respect to our six implemented example mage client tools.


\subsubsection{Automatically Generating Code}
Our nine participants held a mix of experience, as is common with data workers.
Some practitioners had expertise in statistics or ML but were relatively new to programming, while some practitioners were expert programmers with less data/ML experience.
More senior practitioners had expertise in both. A key theme we found was how expertise impacted how practitioners reacted to the idea of a flexible GUI/code tool automatically generating code for them.

Participants who were less confident with programming particularly appreciated the ability of mage tools to generate code for them while allowing them to use more familiar modalities like the data sheet \verb|table|. Some more novice programmers even expressed a preference that generated code be hidden by default, so that they did not even have to look at it. Some notebook environments like Google's Colab\footnote{Google Colaboratory is a web-based notebook environment that allows live multi-user editing \url{https://colab.research.google.com}} do offer code hiding features, where a code cell can be hidden by default but then shown if expanded.
Meanwhile, programming experts tended to reject some tool ideas that they felt they could do much faster by writing code.

An example of participants' divide on code generation is the \verb|save| tool for exporting chart figures to file (Figure~\ref{fig:io}).
Some participants greatly appreciated this functionality as a way to prevent them from having to memorize finicky syntax or make silly mistakes saving their chart incorrectly. However, programming experts said they were unlikely to use a tool such as this when they already know the correct syntax.


\subsubsection{Visual Data Selection}
All participants wanted easier ways to select data and use direct manipulation to pull a selection into code.
This kind of interaction is demonstrated above under ``Multi-tool Use Case'' (Figure~\ref{fig:drag}).
Practitioners value data selection through direct manipulation for multiple purposes, such as being able to interactively label data or interactively choose what data to explore more closely.


\subsubsection{Interact to Explore Model Performance}
Since most participants dealt with developing machine learning (ML) models as part of their work, participants were particularly enthusiastic about having interactive tools to work with to help them understand model performance.
Even with the availability of tests and metrics, understanding model performance is a challenging exploratory task for practitioners, especially for ML novices~\cite{patel2008investigating, yang2018grounding, cabrera2019fairvis}.
Interactive tools can help users better understand model performance~\cite{patel2010gestalt, ren2016squares}, and a benefit of mage is that client tools can take in a user's model data and context, directly within their active code environment, and provide in-the-moment interactive support in situ without switching to an outside GUI tool. An example of this is demonstrated in Figure~\ref{fig:confusion}, with the \texttt{confusion} tool. While the \verb|confusion| tool was rated highest among all nine participants, usually \verb|confusion| prompted a discussion for how participants hoped for such interactive tools to be available for much more complex model analysis tasks. One simple tool idea suggested by a participant was to have an interactive way to explore trade-offs in performance of their model by letting them interactively tune their preferred threshold for precision versus recall. 


\subsubsection{Make Good Practices Easier}

Participants were particularly excited about the ability in \verb|datasplit| to compare distributions, as this is a best-practice that is rarely provided by default. Although of course all functionality in \verb|dataplit| is \textit{possible} in raw code, a GUI makes suggested operations readily apparent to novice users, who may not yet have the ML know-how to follow best-practices without scaffolding. Data and ML programming APIs ideally encapsulate best-practices as the default capabilities they provide to users~\cite{bloch2006design}.
Interactive GUI tools also have more opportunity to nudge users towards best-practices when placed directly into the user's work environment.


\section{Future Work}
Next, we cover areas of future work towards making mage more powerful and flexible for data workers and tool creators:

\subsubsection{Future Work: Generated Code Quality}
During exploratory programming, data workers try out many possibilities that don't always contributing to the final result~\cite{kery2017variolite}.
Simply generating new code for each new user action can unfortunately result in an excess of redundant and obsolete code.
A simple example is shown below:

\begin{figure}[H]
    \centering
    \includegraphics[width=.9\columnwidth]{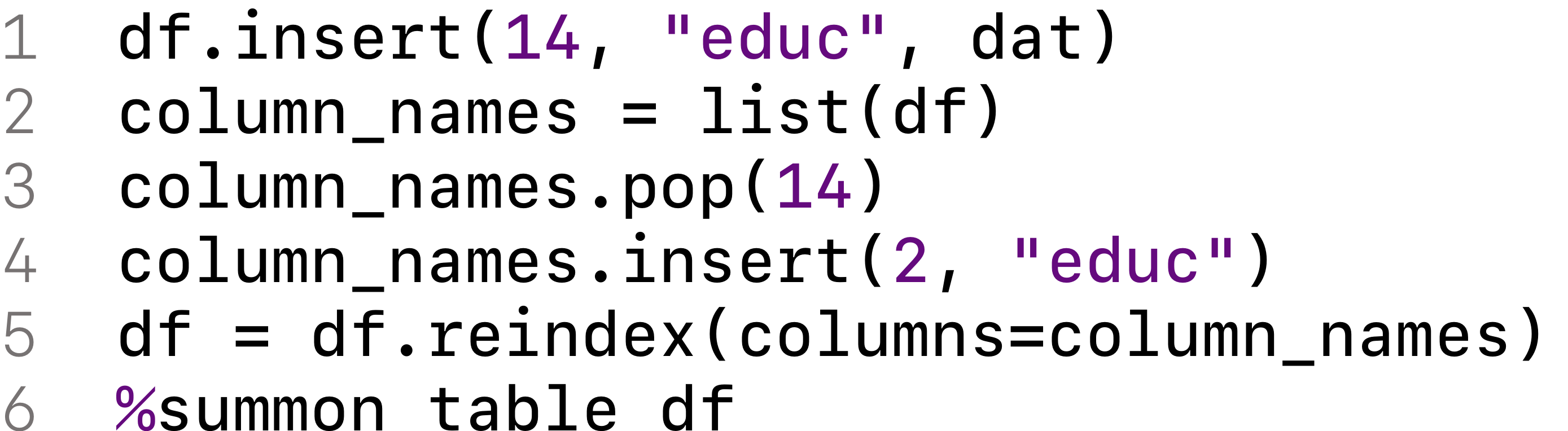}
    \vspace*{-0.3cm} 
    \label{code-bad}
\end{figure}

Here the user inserts a new column at one position, then repositions it.
This sequence should be simplified to just:

\begin{figure}[H]
    \centering
    \includegraphics[width=.9\columnwidth]{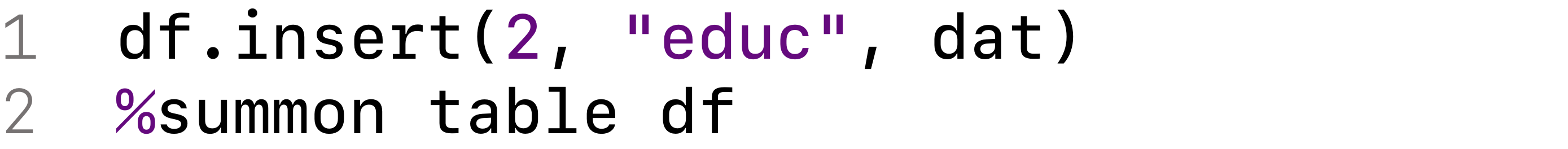}
    \vspace*{-0.3cm} 
    \label{code-good}
\end{figure}

For a single tool, code quality is straightforward to optimize.
For the \texttt{table} tool, we implemented \texttt{table}'s own internal state dictionary object, which is updated by each action in the \texttt{table}.
Instead of adding a new line of code for each action, mage gives the option of fully rewriting the \textit{writable area} of all auto-generated code (unrecognized user code is considered not writable) in the user's code cell.
So for \texttt{table}, after each user action \texttt{table} uses its own state dictionary to output a \textit{list of templates} corresponding just to the final current state of the table.
This results in compact code as above.

Our challenge lies in \textit{generalizing} this kind of optimization to any tool, rather than leaving it to a tool author to implement.
We found that to create a compact internal representation of state, as in \texttt{table}, relies on both having some knowledge of how to operationalize state into a dictionary and order of operations, both of which are specific to the tool domain.
Improving this process in an area for future work.

\subsubsection{Future Work: Movement Between GUI Tools}
Although we prototyped some ways that a user can move data between two GUI tools (see the section on Multi-tool Use Case above), many questions remain open about how a user might transfer content between two GUI tools, and not just between GUI and code.
A simple use case for this may be to create a set of coordinated visualizations, like an impromptu data dashboard, that all update based on data selection from a table.
Or, similarly, a data worker may want several GUI tools designed for providing model metrics to display metrics specific to different data slices within the model or different tradeoffs between precision and recall.
The chief technical challenge in coordinating multiple GUI tools in such a way is designing an easy-to-use protocol by which a tool creator can specify what kinds of \textit{other tools} and other data signals their tool can work with.
This adds extra complexity for tool builders, which would need careful design to ensure the process was as lightweight for creators as possible.

\hl{More generally an overall challenge for mage, or any future toolkit like it, is to make the process for individual tool builders easier. Despite providing tool builders a bulk of analysis and code generation to enable fluid GUI/code work, our work demonstrates a number of complications where tool-specific understanding is currently needed from tool builders.}

\section{Conclusion}
In this work, we have discussed mage, an API for allowing smooth transitions between GUI and code work in notebooks, as well as new kinds of tool avenues this GUI/code back-and-forth enables, demonstrated through six example tools.
Our hope is that this work can enable and inspire new support tools to give data workers flexible interactive help at the moment as they need it.
However, the core idea of flexible transitions between tools is much bigger than our work here, or computational notebooks.
Seamless transitions between different devices in a user's environment or seamless transition between code and visuals in 3D graphics programs are both common examples of this broader technique.
\ham{I know that we try to claim generalizability here, but I think it's a bit too stretching given that we haven't give enough justification why seamless transition is equally or more important in 3D.  Seamless transition between devices is also a quite different story.}
For data work, blending modalities of different kinds of tools practitioners use offers a new interaction space for richer interactions and more context-aware support tailored to a practitioners' specific workflow.
We hope to see more HCI systems work in these intersections to support data workers in coming years.

\section{Acknowledgements}
We thank all of our study participants, and everyone who provided feedback on this work, including: Prof. Brad A. Myers at Carnegie Mellon University, all of our reviewers, and our colleagues at Apple Inc.

%
%
%
%
%

\balance{}

\bibliographystyle{SIGCHI-Reference-Format}
\bibliography{paper}

\end{document}